\documentclass[pra,amsmath,showkeys,showpacs]{revtex4}
\usepackage{graphicx}
\newcommand{\Mew}[2]{\mathsf{E}\kern-1pt\llap{$\vert$}(#1;#2)}
\newcommand{\bra}[1]{\langle{#1}|}
\newcommand{\ket}[1]{|{#1}\rangle}
\newcommand{\braV}[1]{\bra{\vec{#1}}}
\newcommand{\ketV}[1]{\ket{\vec{#1}}}
\newcommand{\OpUnit}{\openone}
\newcommand{\RR}{R}
\newcommand{\Trace}{\text{Tr}\,}
\newcommand{\Prob}{\text{Prob}\,}
\newcommand{\Probc}{\text{Probc}\,}

\newcommand{\Wp}{\text{wp}}
\begin{document}
\title{Time operator within projection evolution model}
\author{Andrzej~G\'o\'zd\'z}
\email{gozdz@neuron.umcs.lublin.pl}
\author{Mariusz~D\c{e}bicki}
\email{mdebicki@tytan.umcs.lublin.pl}
\affiliation{Institute of Physics, University of Marie~Curie--Sk\l{}odowska,\\
	     pl.\ M.~Sk\l{}odowskiej--Curie 1, 20--031 Lublin, Poland.}
\date{\today}
\begin{abstract}
We apply the projection evolution approach to the particle  detection process
and calculation of the detection moment. Influence of the essential system
properties on the evolution process is discussed. It is shown, that  using 
only the projection postulate  in the evolution scheme allows to understand the
time  as a kind of observable.
\end{abstract}

\maketitle

\section{Introduction}
\label{sec:introduction}
The usual formulation of quantum mechanics leads to many conceptual problems. 
Most of the proposed interpretations concerning the measurement process is 
unsatisfactory. The great interest in quantum computers and wide range of
obstacles to them, caused mainly  by decoherence of the entangled states has
contributed to retrospect the essentials  of the quantum mechanics.

The main problem is that there are two quite different evolution laws.  The
Schr\"odinger equation describes the evolution of the quantum objects as long
as  it is not disturbed by the experiment. It is unitary and completely
deterministic process. On the other hand there is the projection postulate
which operates whenever the object is  affected by the measurement. Here one
can calculate only the probabilities for each possible outcome in the
experiment,  but the change of the object state is unpredictable.  Moreover,
there is no way to estimate the moment of time when the process occurs.

A measurement of time stands as the separate general problem of the quantum
mechanics. Time has been treated as a parameter in the evolution law for
decades,  even it has been proven that the time operator can not be defined
properly within standard quantum mechanics.

The Pauli's theorem \cite{Paul58} states that the self-adjoint time operator
implies an unbound  continuous energy spectrum. That means, it is impossible to
build a self-adjoint time operator canonically conjugate to a Hamiltonian
bounded from below.
However, some attempts to apply the concept of time as an observable were made, 
according to the statement, that any given observable  uniquely characterized
by the probability distribution of the measurement results in the different
states accessible to  the system.

Thus there were proposed several types of times corresponding to some types of
measurements, e.g. the time of arrival introduced by Aharonov and Bohm
\cite{Ahar61}, further investigated by e.g. \cite{Busc94, Busc95, Egus99,
Muga98}, also \cite{Allc69, Wern86}, the tunneling time or the time of a
quantum clock given by a phase variable.

Another concept of time operator was introduced by Olkhovsky, Recami and others
\cite{Olkh74}, where the basic idea is to extract it from the average "presence
time" relation  $\langle t\rangle_{x=0}$. More recent approach applies a
positive operator valued measure (POVM) \cite{Busc94, Holev82, Gian96} to this
problem. Some other publications presents the time-of-arrival problem using
some kind of "screen observables" \cite{Wern86}. More detailed review can be
found in e.g. \cite{Kuus01}. An interesting idea is also the Event-Enhanced
Quantum Theory which replaces the Schr\"odinger equation by a special
deterministic algorithm \cite{Blan98}, however it seems to be unsatisfactory
introducing some "non quantum" elements.

But apart from the pure time operator the problem lies in the process of time
measurement and evolution. The first model proposed by Allcock \cite{Allc69}
consisted of a free particle and an interacting Hamiltonian in the form of the
complex potential. The solutions, however, were in disagreement with the
Heisenberg uncertainty relation. Aharonov and Bohm proposed to consider a
system consisting of "a clock" among other quantum objects (particles,
apparatus, etc.). The time of interaction is thus determined by a physical
observable of this clock particle \cite{Ahar61} where the corresponding time
operator was obtained by a simple symmetrization of the classical expression
$t=mx/p_x$.

But what is that mysterious force described by a complex potential or "a
clock"? What is its foundation? And what defines the physical process as an
experiment which obeys the projection postulate rather than the unitary
transformation? The questions are still open until today.

Some recent work was made concerning quantum computation process where the
evolution of a qubit is assumed to be the process of a succession of some
measurements, e.g. as the result of  some additional periodically interrupting
influence \cite{Audr02}. Very close model was  introduced by Nielsen
\cite{Niel01}, who showed that no coherent unitary dynamics is needed at  all
in order to simulate the quantum computer. Thus projective measurements are
universal for quantum computation \cite{Luen01, Fenn01}.

We have made one step forward following the hypothesis of the projection
evolution \cite{Gozd03},  which is a sequence of some measurements made
permanently by the Nature. Each system behaves like it was affected by some
kind of the "apparatus" but we postulate it is its natural inner property. A
set of projection evolution operators responsible for the measurement at each
moment of evolution depends on the essential properties of the system under
consideration. Detailed description of the approach can be found in 
\cite{Gozd03}.

Within this idea many of the mentioned above problems is solved, in
particular the Schr\"odinger equation is a special case of the projection
evolution. Moreover, within this idea,  there is no need to introduce an
observer  because the "collapse of states" is processed due to the
postulated fundamental Law of the Nature. Further, this model lets to introduce
a kind of the time operator, which can not be done within the  standard quantum
mechanics. More detailed information about the time observable is presented in
the next  section of this paper.

\section{The time operator}
\label{sec:time_operator}
The projection evolution model as described in \cite{Gozd03} defines the
evolution as a process  of causally related physical events, ordered by the
evolution parameter $\tau$, called in \cite{Gozd03} the etime.  The time-like
variable parameter $\tau$ has been already regarded by Caster and Reznik
\cite{Cash99} as the state of the clock particle within the time-of-arrival
problem, but in contradiction to that hypothesis, within the projection
evolution approach each of the events is related to a measurement "made",
according to the proposed procedure, by the Nature.  That means, the state of a
physical system at the moment (step of evolution) denoted by $\tau$ is the
projection  of the previous state with respect to the essential system
properties, as it is shown below (for simplicity we assume here a discrete
structure of the evolution parameter (etime) $\tau$):
\begin{equation} \label{eq:II.1}
\rho(\tau_{n+1};\nu_0,\nu_1,\dots,\nu_{n+1})=
	\frac{\Mew{\tau_{n+1}}{\nu_{n+1}}\rho(\tau_n;\nu_0,\nu_1,\dots,\nu_n)
		\Mew{\tau_{n+1}}{\nu_{n+1}}}
	{\Trace[\,\Mew{\tau_{n+1}}{\nu_{n+1}}
	         \rho(\tau_n;\nu_0,\nu_1,\dots,\nu_n)
		\Mew{\tau_{n+1}}{\nu_{n+1}}\,]}.
\end{equation}
Here $\rho(\tau_n;\nu_0,\nu_1,\dots,\nu_n)$ is the quantum density operator
describing the state of the physical system at n--th step of the evolution. 
The next step is specified  by the projection evolution  operator
$\Mew{\tau_{n+1}}{\nu_{n+1}}$ randomly chosen from a set of possible
projections which constitute the orthogonal resolution of unity, each
representing the properties of our physical system at the moment $\tau_{n+1}$
of the etime. The orthogonal resolution of unity, for discrete quantum numbers
$\nu$ are defined by the conditions:
\begin{eqnarray} \label{eq:II.1a}
&\Mew{\tau}{\nu}\Mew{\tau}{\nu'}=\delta_{\nu\nu'}\Mew{\tau}{\nu} \nonumber \\
&\sum_\nu \Mew{\tau}{\nu} = \OpUnit.
\end{eqnarray}
and can be naturally generalized to arbitrary sets of quantum numbers. 

The probability distribution of choosing the specified projection is given by
a rather standard  formula:
\begin{equation} \label{eq:II.2}
\Prob(\tau_{n+1};\nu_0,\nu_1,\dots,\nu_{n+1})=
\Trace[\,\Mew{\tau_{n+1}}{\nu_{n+1}}\rho(\tau_n;\nu_0,\nu_1,\dots,\nu_{n})
	\Mew{\tau_{n+1}}{\nu_{n+1}}\,].
\end{equation}
This is, in principle, the conditional probability because the 
eq.(\ref{eq:II.2}) describes a probability of choosing next state, under
assumption that the system is already in the given state 
$\rho(\tau_n;\nu_0,\nu_1,\dots,\nu_{n})$.

It is shown in \cite{Gozd03} that assuming the projection evolution operator in
the form of resolution of unity shifted by an unitary operators like traditional
unitary evolution operator for closed systems:
\begin{equation} \label{eq:II.2b}
\Mew{\tau}{\nu}=U(\tau-\tau_0)\Mew{\tau_0}{\nu}U^\dagger(\tau-\tau_0),
\end{equation}
under some assumptions, such process while considering the continuous
measurements leads to the unitary evolution of the Schr\"odinger's type 
\begin{math}
\rho(\tau)=U(\tau-\tau_0)\rho'(\tau_0)U^\dagger(\tau-\tau_0),
\end{math}
where $\rho'(\tau_0)$ is the state chosen by the Nature just before
$\tau_0$.
One needs observe here that this approach allows to consider time on nearly the
same foot as any other quantum observable because the time should be related to
the parameter etime ordering causal events, but it is formally independent of
it. 

The projection evolution procedure suggest also a possibility to consider two
types of observables. 

The first type consists of  physical observables related to real interactions
with the system under considerations and reducing its state due to the
interaction. This kind of observables can be a part of the projection evolution
operator.

The second type of observables (information observables) are more theoretical
ones because they allow only to investigate structure of states.

Let the decomposition of unity  $\{M_A(a)\}$ be an information observable and
$\rho$ be a state of the system. In this case the expression
$\Trace(M_A(a)\rho)$ should  rather be interpreted as a "potential probability"
which give us an information about the structure of the system state but not
affect it.  Such information should not have any influence into the real
process of the evolution.

On the other hand, it is important to notice that probably most of possible
observables can be used formally in both meanings. \\
However, an existence of the first type observables, for a given physical
system, is determined by structure of the system and its interactions.

Assuming now a nonrelativistic four dimensional spacetime $x=(x_0=t,\vec{x})
\in R^4$, using the specified reference frame, there is now the possibility to
build a well defined "time operator" $M_T(t)$ projecting onto the subspace of
the simultaneous events.  As mentioned in the previous section that is
impossible in the standard quantum mechanics where time is a parameter. For this
simple model of space and time the states responsible for the event "to be in
$\vec{x}$ at the time $t=x_0$" can be described by the Dirac $\delta$-type
distribution $\delta^4(y-x)$ corresponding to the appropriate state $\ket{x}$.
Making use this notation the appropriate projection operator (generalized
resolution of unity) can be written as:
\begin{equation} \label{eq:II.2a}
M_T(t)=\int_{R^3}\, d^3\vec{x}\, \ket{x}\bra{x},
\end{equation}
where $d^3\vec{x}$ denotes an element of 3 dimensional volume of the coordinate
space. 

This operator should give us all information about an overlap of our system
over the subspace of states corresponding to any position at time $t$. We
suppose also that the operator can be used in non-relativistic case (it is not
covariant in respect to the Lorentz transformations) as an approximation of the
physical observable which detect objects at a given time $t$. In this case it
must be related to a specific physical structure, like a detector, which is
able to interact with our system  at this given time.

Within the projection evolution model, as was already mentioned above, the
whole process of evolution is ordered by the special  evolution
parameter~$\tau$. Thus the sequence of all events is independent from the
observer's configuration space. Using the time projection operator we can now
determine the potential probability density, that the state  described by
$\rho(\tau)$ can be found at the specified moment $t$, as follows:
\begin{equation} \label{eq:II.3}
	\Prob(t;\rho(\tau))=\Trace(M_T(t)\rho(\tau)M_T(t)).
\end{equation}
As mentioned previously the operator (\ref{eq:II.2a}) can be also used as a
kind of "time trigger". Then every event enumerated with the $\tau$ parameter
can be related with the real world, and can occur only at the specified moments
of time with well defined probability.

In this way we can describe the system evolution with respect to the evolution
parameter $\tau$ as well as the time $t$. 

The following example of the particle detection presented in the next sections
shows  how we can describe the evolution, its duration and the moment of the
measurement using  only the projection evolution scheme.

\section{Measurement in the projection evolution scheme}
\label{sec:measurement}
Let us consider a closed system of a single particle "moving" towards a
measurement device. To simplify considerations let us assume the device is a
kind of detector that registers  a particle when it comes across a definite
region of the space. Physical mechanisms related to the process of absorption
and detection are beyond of our scope.

Due to the projection evolution model we need to construct an appropriate set
of projection  operators $\Mew{\tau}{\nu}$, responsible for time evolution. 
Here $\tau$ is a real c-number evolution parameter that enumerates the 
subsequent evolution steps, and $\nu$ represents any set of quantum  numbers
describing the system properties.

Because the particle position and its linear momentum (due to the Heisenberg
uncertainty principle) cannot be well determined simultaneously,  we assume, in
general, that the free particle can evolve in a form of some wave packets
$\ket{\nu}$ which constitute the orthonormal basis within the state space of
the particle, as written below:
\begin{equation} \label{eq:III.1}
\ket{\nu} = \int_{\RR^4} d^4 k\ \alpha_\nu (k)\;\ket{k},
\end{equation}
where $k \equiv (k_0,\vec k)$ denotes a wave vector  with $k_0$ proportional to
the particle energy  and $\vec k$ represents its linear momentum. The vector
$k$ can be thought  as kind of energy--momentum four-vector in $R^4$.\\
The shape of the wave packet is fully determined by $\alpha_\nu(k)$ 
that should fulfill the normalization condition 
\begin{math}
\int_{\RR^4} d^4 k\ \vert\alpha_\nu (k)\vert^2=1.
\end{math}	
That means the free particle can be in one of the possible states $\ket{\nu}$
at every moment (step) of its evolution.

Our considerations allow to  define the family of projections which are an
orthogonal resolution of unity and can determine a motion of free particle
within the projection evolution approach 
\begin{equation} \label{eq:III.2}
M_\Wp(\nu) = \ket{\nu}\bra{\nu}.
\end{equation}
In general $\nu$ are related to some system (particle) properties under
consideration. \\
In particular the free particle evolution, one can expect,  could be described
by a set of Gauss-like wave packets or the coherent-like states in order to
minimize the Heisenberg uncertainty principle.

Another set of projection operators we have to define, is related to the
detector as a measurement device. In the simplest case the appropriate
operators can be written as follows:
\begin{equation} \label{eq:III.3}
M_D(\Delta)=\int_\Delta d^3 \vec{x}\ \ketV{x}\braV{x},
\end{equation}
where $\ketV{x}$ denotes the generalized eigenvectors of the position operators
and  $\Delta$ describes the detector shape, that is the region where a particle
is registered  when arriving. This operator "projects" onto the coordinate
states belonging  to the detector. The complementary operator 
\begin{math}
M_D' \equiv M_D(\nu = 0) = \OpUnit - M_D(\nu = \Delta)
\end{math}
takes care of the situation when the particle is outside the detection area.

First we consider the evolution with the detector  without the "time trigger".
In the following we take into account the wave packets which are
$\delta$--localized in time and spread out over the coordinate space.

Under this assumption we can construct the evolution operators as follows:
\begin{equation} \label{eq:III.4}
\Mew{\tau}{\nu} = \begin{cases}
U(\tau-\tau_0)\,M_\Wp(\nu)\,U^\dagger(\tau-\tau_0); & 
	\text{for $\tau_0 \leq \tau < \tau_D$ and $\nu$ of the form 
	(\ref{eq:III.1})}, \\
M_D(\nu); & 
	\text{for $\tau = \tau_D$ and $\nu = \Delta$ or $\nu = 0$}, 
\end{cases}
\end{equation}
where $U(\tau)$ is the unitary evolution operator of the Schr\"odinger like
form,  very close to the traditional one for free particle, generated by the
kinetic energy: 
\begin{equation} \label{eq:III.5}
U(\tau) = e^{i\beta_0 \hat{k}^0\tau}e^{-i\beta\hat{\vec{k}}^2\tau},
\end{equation}
where  $\beta_0$ and $\beta$ are some coefficients dependent on physical
system and $\hat{k}^0,\hat{\vec{k}}$ denotes here the "four-momentum operator"
for which the vectors $\vert k\rangle$ are generalized eigenvectors i.e.,
 $\hat{k}^\mu \ket{k} = k^\mu \ket{k}$ \cite{Gozd03}.

According to the projection evolution model \cite{Gozd03}, the system can
follow  any path of the evolution. In particular it can travel as a free
particle  from a source in the form of wave packet (\ref{eq:III.1}) denoted by
$\ket{\mu}$, and after the specified evolution steps $\tau_D$ hit the
detector. It can also miss the detector of course, but this is less interesting
now. The probability of the first situation can be calculated using the
formula (\ref{eq:II.2}):
\begin{equation} \label{eq:III.6}
\Prob (\tau = \tau_D; \nu_1 = \mu, \nu_2 = \Delta,t) =
\Trace [\,M_D(\Delta)\,U(\tau_D-\tau_0)\,M_\Wp(\mu)\,
\rho_0\, 
M_\Wp(\mu)\,U^\dagger(\tau_D-\tau_0)\,M_D(\Delta)\,], 
\end{equation}
where $\rho_0$ is a quantum density operator describing the particle initial
state.  It depends on the source properties and physical mechanisms associated
with the  particle creation process.

Using as a basis the generalized eigenstates $\ket{x}=\ket{x^0,\vec x\,}$ of
formal  positions operators within the spacetime $R^4$,   together with
(\ref{eq:III.2})  and (\ref{eq:III.3}) we can rewrite the expression for the
probability (\ref{eq:III.6}) as:
\begin{equation}\label{eq:III.7}
\begin{split}
\Prob (\tau = \tau_D; \nu_1 = \mu, \nu_2 = \Delta) &= 
	\int_{\RR^4} d^4 x\ \bra{x} 
		\int_\Delta d^3 \vec{x}\,'\ \ket{\vec{x}\,'}\bra{\vec{x}\,'}\,
  U(\tau_D - \tau_0)\,\ket{\mu}\bra{\mu}\rho_0\ket{\mu}\\ 
  \bra{\mu}\,U^\dagger(\tau_D - \tau_0)&\int_\Delta d^3 \vec{x}\,''
	\ \ket{\vec{x}\,''}\bra{\vec{x}\,''}x\rangle	 =
\ \bra{\mu}\rho_0\ket{\mu}\int_\RR dx^0\int_\Delta d^3 \vec{x}
		\ \Big|\bra{x}\,U(\tau_D - \tau_0)\,\ket{\mu}\Big|^2
\end{split}
\end{equation}
As one can see the first part of (\ref{eq:III.7}), that is  
$\bra{\mu}\rho_0\ket{\mu}$ is the probability distribution of finding the
particle in the state $\ket{\mu}$. In particular, it can be equal 1 if only the
initial state created from the source is just  a pure vector $\ket{\mu}$.

More interesting is the second part of (\ref{eq:III.7}) describing the
probability of finding the particle within the detector after the specified
steps of evolution (\ref{eq:III.4}).
Let us denote this probability by:
\begin{equation}\label{eq:III.8}
\pi_D(\nu) \equiv \int_\RR dx^0\int_\Delta d^3 \vec{x}
	\ \Big|\bra{x}\,U(\tau_D - \tau_0)\,\ket{\mu}\Big|^2 = 
\int_\RR dx^0\int_\Delta d^3 \vec{x}\ \left|\,\int_{R^4} d^4 k 
\ \bra{x}\,U(\tau_D - \tau_0)\,\alpha_{\mu}(k)\,\ket{k} \right|^{\;2}
\end{equation}
The statement given above becomes much simpler if we consider the states for
which the coefficients $\alpha_\nu (k)$ can be factorized into  the energy
(time) and linear momentum (space) parts. 
Thus, we assume,  the wave packet  $\ket{\mu}$ defined by (\ref{eq:III.1})
fulfills the following condition:
\begin{equation}\label{eq:III.9}
\alpha_{\mu}(k) \equiv \kappa_{\mu}(k^0)\;
\alpha_{\mu}(\vec k),
\end{equation}
with the appropriate normalization 
\begin{math}
\int_{\RR} dk^0\ \vert\kappa_\mu (k^0)\vert^2=1
\end{math}	
and 
\begin{math}
\int_{\RR^3} d\vec k^3\ \vert\alpha_\mu (\vec k)\vert^2=1.
\end{math}	
\\
This is the particular case that strictly corresponds to the non-relativistic 
description of the common quantum mechanics when time is treated as the
parameter. 
  
According to  (\ref{eq:III.9}) we can now rewrite (\ref{eq:III.8}) 
as follows:
\begin{equation}\label{eq:III.10}
\pi_D(\mu) = 
\int_\RR dx^0\ \left|\,\int_\RR dk^0
 \ \bra{x^0}\,e^{i\beta_0 k^0\tau_D}\kappa_\mu(k^0)\,\ket{k^0}\right|^{\;2}
\int_\Delta d^3 \vec{x}\ \left|\,\int_{R^3}d^3\vec k 
 \	\bra{\vec x}\,e^{-i\beta\vec k^2\tau_D}
	\;\alpha_{\mu}(\vec k)\,\ket{\vec k} \right|^{\;2}
\end{equation}
The mentioned separation of the time and space coordinates makes 
further calculations easier to process. Moreover, making use of the 
orthogonal resolution of unity $\int_\RR dx^0\ \ket{x^0}\bra{x^0} = 1$
it is easy to show that the time dependent part of the (\ref{eq:III.10}) 
is equal to one:
\begin{displaymath}
\int_\RR dx^0\ \left|\,\int_\RR dk^0
 \ \bra{x^0}\,e^{i\beta_0 k^0\tau_D}\kappa_\mu(k^0)\,\ket{k^0}\right|^{\;2} = 1,
\end{displaymath}
which means, it does not affect anything in the process (measurement)
probability.
 
This is the case when the system properties do not change their character 
during the evolution process. For example, if the detector could register 
the particles only for the specified period, the time dependent factor 
would be significant in the probability result. 

We assumed, here,  the detector is a kind of stationary device which register a
particle  immediately after it reaches the detection area.

Describing the particle evolution we are interested if it is able to come
across the detector area. For this purpose we specify the state $\ket{\mu}$ as 
the packet of states $\ket{k}$ equally distributed between $\vec k-\vec{\Delta
k}$ and  $\vec k+\vec{\Delta k}$ as defined by (\ref{eq:III.1})  with
(\ref{eq:III.9}), where
\begin{equation}\label{eq:III:10a}
\kappa_{\mu}(\vec k)=\begin{cases}
 \alpha = \text{const}; & 
  \text{for $\vec k \in (\vec k_0-\vec{\Delta k},\vec k_0+\vec{\Delta k})$}\\
  0; & \text{for $\vec k \not\in (\vec k_0-\vec{\Delta k},\vec k_0+
       \vec{\Delta k})$}
\end{cases}
\end{equation}
and $\vec k_0$ is here a constant vector of linear momentum centering the wave
packet.

Using the normalization condition $\bra{\mu}\mu\rangle = 1$ we
can calculate 
\begin{math}
\alpha=(1/{2\Delta k_x})^{1/2}(1/{2\Delta k_y})^{1/2}(1/{2\Delta k_z})^{1/2}.
\end{math}
Thus the linear momentum of the particle, as it can be shown in a simple way,
is represented by a spectrum  of values between $\hbar(\vec k_0-\vec{\Delta
k})$ and $\hbar(\vec k_0+\vec{\Delta k})$ with the average value equal
$\hbar\vec k_0$. 

Now after the reduction of the time coordinate integrals 
and using the specified form of the wave packet (\ref{eq:III:10a})
we can rewrite (\ref{eq:III.10}) as follows:
\begin{equation}\label{eq:III.11}
\begin{split}
\pi_D(\mu)&=\alpha^2\int_\Delta d^3\vec{x}\ \left|\,
\int_{\vec k_0-\vec{\Delta k}}^{\vec k_0+\vec{\Delta k}}d^3\vec k
\ \braV{x}\,\vec k\rangle\ e^{-i\beta\vec k^2\tau_D}\right|^{\,2}\\
&=\alpha^2\left(\frac{1}{2\pi}\right)^3\int_\Delta d^3\vec{x}\ \left|\,
\int_{\vec k_0-\vec{\Delta k}}^{\vec k_0+\vec{\Delta k}}d^3\vec k
\ \,e^{i\vec k\vec{x}}\;e^{-i\beta\vec k^2\tau_D}
\right|^{\,2}
\end{split}
\end{equation}
According to the regarding system properties we are tending to minimize the
uncertainty of the particle location and its linear momentum but both strongly
depend on the wave packet width.

Let us consider an approximation which
corresponds to the small linear momentum spread around $\vec{k_0}$. Using the
linear expansion to the first order in the small deviations of $\vec{k}$ from 
$\vec{k}_0$ we have
below:
\begin{equation}\label{eq:III.12}
\vec k^2=(\vec{k}_0+\delta\vec{k})^2\approx 2\vec{k}\vec{k}_0- {\vec{k}_0}^2
\end{equation}

This approximation prevents us from using any numerical receipts in order to
calculate the inner integral of (\ref{eq:III.11}), which is presented below:
\begin{equation}\label{eq:III.13}
\begin{split}
\int_{\vec k_0-\vec{\Delta k}}^{\vec k_0+\vec{\Delta k}}d^3\vec k
\ \,e^{i\vec k\vec{x}}\;e^{-i\beta\vec k^2\tau_D}
&=e^{-i\beta\vec k_0^2\tau_D}
\int_{\vec k_0-\vec{\Delta k}}^{\vec k_0+\vec{\Delta k}}d^3\vec k
\ \,e^{i\vec k\left(\vec{x}-2\beta\vec k_0\tau_D\right)}
\\=e^{-i\beta\vec k_0^2\tau_D}
e^{i\left(\vec{x}-2\beta\vec k_0\tau_D\right)\vec k_0}
&\left[\frac{2\sin{[(x-2\beta\vec k_{0x}\tau_D)\Delta k_x]}}%
            {x-2\beta\vec k_{0x}\tau_D}
\ \frac{2\sin{[(y-2\beta\vec k_{0y}\tau_D)\Delta k_y]}}%
       {y-2\beta\vec k_{0y}\tau_D}
\ \frac{2\sin{[(z-2\beta\vec k_{0z}\tau_D)\Delta k_z]}}%
       {z-2\beta\vec k_{0z}\tau_D}
\right]\\
\end{split}
\end{equation}\\

The second integral of the (\ref{eq:III.11}) depends on the shape of the
detector. The rectangular box seems to be  the simplest one, so let us choose
that shape. The probability of detection of the particle can be now calculated
from (\ref{eq:III.11}) with respect to the previous assumptions
(\ref{eq:III.12}), as follows:
\begin{equation}\label{eq:III.14}
\begin{split}
\pi_D(\mu)&=\alpha^2\Big(\frac{1}{2\pi}\Big)^3\! \!
\int^{a_2}_{a_1}\llap{\text d}x\!\!
\int^{b_2}_{b_1}\llap{\text d}y\!\!
\int^{c_2}_{c_1}\llap{\text d}z\Big| \ e^{-i\beta\vec k_0^2\tau_D}
e^{i(\vec x-2\beta\vec k_0\tau_D)k_0}			\\ & \qquad
\frac{2\sin[(x-2\beta\vec k_{0x}\tau_D)\Delta k_x]}%
     {x-2\beta\vec k_{0x}\tau_D} 
\frac{2\sin[(y-2\beta\vec k_{0y}\tau_D)\Delta k_y]}%
     {y-2\beta\vec k_{0y}\tau_D}
\frac{2\sin[(z-2\beta\vec k_{0z}\tau_D)\Delta k_z]}%
     {z-2\beta\vec k_{0z}\tau_D}  
\Big|^2 =\\
\frac{1}{\pi\Delta k_x}
\int^{a_2}_{a_1}&\text d{x}
\frac{\sin^2[(x-2\beta\vec k_{0x}\tau_D)\Delta k_{x}]}%
     {(x-2\beta\vec k_{0x}\tau_D)^2}
\frac{1}{\pi\Delta k_y}
\int^{b_2}_{b_1}\text d{y}
\frac{\sin^2[(y-2\beta\vec k_{0y}\tau_D)\Delta k_{y}]}%
{(y-2\beta\vec k_{0y}\tau_D)^2}
\frac{1}{\pi\Delta k_z}
\int^{c_2}_{c_1}\text d{z}
\frac{\sin^2[(z-2\beta\vec k_{0z}\tau_D)\Delta k_{z}]}%
{(z-2\beta\vec k_{0z}\tau_D)^2}
\end{split}
\end{equation}
Here $(a_1,a_2),\,(b_1,b_2),\,(c_1,c_2)$ denote the measurement device corner
points on the defined reference frame.

As one can see the integral is easily separable into three one-dimensional
integrals,  each of them representing the probability of the particle detection
in the $k_x, k_y, k_z$ direction, respectively.\\
After a bit of algebra we can derive the following:
\begin{equation}\label{eq:III.15}
\begin{split}
\pi_D(\mu_x)\equiv\frac{1}{\pi\Delta k_x}\int_{a_1}^{a_2}&\text dx
\frac{\sin^2[(x-{v_{gx}}\tau_D)\Delta k_x]}{({x}-{v_{gx}}\tau_D)}=
\frac{\sin^2[(a_1-{v_{gx}}\tau_D)\Delta k_x]}%
     {\pi({a_1}-{v_{gx}}\tau_D)\Delta k_x}
-\frac{\sin^2[(a_2-{v_{gx}}\tau_D)\Delta k_x]}%
      {\pi({a_2}-{v_{gx}}\tau_D)\Delta k_x}\\ 
&-\frac{1}{\pi}\Big\{
\text{si} [2\Delta k_x(a_1-{v_{gx}}\tau_D)]-
\text{si}[2\Delta k_x ({a_2}-v_{gx}{\tau_D})]\Big\}
\end{split}
\end{equation}
where $\vec{v}_g\equiv 2\beta\vec k_0$ can be understood as the wave packet
group velocity, well defined if only $\Delta k \ll k_0$ according to the
assumption (\ref{eq:III.12}).

Some example results of the above are presented in the section
\ref{sec:results}.

\section{Time of the measurement}
\label{sec:time_measurement}
The results we have obtained above give us the information about the
probability of the particle detection after the specified number of the
evolution steps $\tau_D$.  
But how we can bound this parameter with time? Some thoughts on this problem
have been already presented in section {\ref{sec:time_operator}}, and now we
are to make some calculations using the time operator $M_T(t)$ defined in
(\ref{eq:II.2a}).

The time operator can be used here either as an information observable allowing
to investigate the temporal structure of states or as a kind of "time
trigger",  an "additional part" of the detector which is responsible for
counting particles at given time. In both cases the calculated probabilities
will be different.

In the first case  the state of the particle entering the detector region
is given by
\begin{equation} \label{eq:IV.2}
\rho(\tau_D;\nu_1=\mu,\nu_2=\Delta) = 
	\frac{M_D(\Delta)\,U(\tau_D)\,M_\Wp(\mu)\,\rho_0\, 
			M_\Wp(\mu)\,U^\dagger(\tau_D)\,M_D(\Delta)}
		{\Trace [\,M_D(\Delta)\,U(\tau_D)\,M_\Wp(\mu)\,\rho_0\, 
			M_\Wp(\mu)\,U^\dagger(\tau_D)\,M_D(\Delta)\,]}
\end{equation}
Using now the same analysis as presented in sec.~{\ref{sec:measurement}} and
after some rather simple algebra, one can find that the equation
(\ref{eq:II.3}) can be rewritten as:
\begin{equation}\label{eq:IV.3}
\Probc(t;\tau_D;\mu,\Delta) \equiv
\Trace [M_T(t)\rho(\tau_D;\nu_1=\mu,\nu_2=\Delta)] = \frac{
	\int_\Delta d^3 \vec{x}\ \left|\,\bra{t,\vec{x}}\,U(\tau_D)\,
	\ket{\mu} \right|^{\;2}
}{
	\int_\RR dx^0
	\int_\Delta d^3 \vec{x}\ \left|\,\bra{x}\,U(\tau_D)\,
	\ket{\mu} \right|^{\;2}
}
\end{equation}
where $\Probc(t;\tau_D;\mu,\Delta)$ can be interpreted as conditional
probability (or probability density) of registration of our particle in the 
detector at time $t$ when the particle is already in the state (\ref{eq:IV.2}).

Using (\ref{eq:III.5}) and taking into account the form of the coefficients
(\ref{eq:III.9}), and after separating the space and time dependent factors the
trace (\ref{eq:IV.3}) can be further rewritten as:
\begin{equation}\label{eq:IV.4}
\Trace [M_T(t)\rho(\tau_D;\nu_1=\mu,\nu_2=\Delta)] = \frac{
  \left|\,\bra{t}\,e^{i\beta_0 \hat{k}^0\tau_D}\,\ket{\mu^0} \right|^{\;2}
}{
  \int_\RR dx^0	
  \left|\,\bra{x^0}\,e^{i\beta_0 \hat{k}^0\tau_D}\,\ket{\mu^0} \right|^{\;2}
}
\end{equation}
Here $\ket{\mu^0} = \int_\RR dk^0\kappa_\mu(k^0)\ket{k^0}$ 
represents the time coordinate part of the particle wave packet. 

It is easy to show that the denominator of the (\ref{eq:IV.4}) 
is equal $1$, so finally we get:
\begin{equation}\label{eq:IV.5}
\Probc(t;\tau_D;\mu,\Delta)=
  \left|\,\bra{t}\,e^{i\beta_0 \hat{k}^0\tau_D}\,\ket{\mu^0} \right|^{\;2}
\end{equation}

The calculation of this conditional probability lets us to predict the
evolution  in terms of the time corresponding to the specified observer and his
reference frame. Thus we will able to verify if our assumptions are correct. In
particular the free particle evolution process should be very close to the
Schr\"odinger's one and in classical limits it should represent any observed or
measured quantity values. Time of the detection of the particle moving towards
the detector is one of these quantities.

The physical conditions require the probability (\ref{eq:IV.5}) should be a
function with well  pronounced maximum, i.e. the function well localized in
time.
For this purpose the amplitudes $\kappa_\mu(k^0)$ in the vector $\ket{\mu^0}$
should be nearly a "plain-wave" type function, because one can show that the 
matrix element:
\begin{equation}\label{eq:IV.6}
	\,\bra{t}\,e^{i\beta_0 \hat{k}^0\tau_D}\,\ket{t_0} = 
\delta (t_0+\beta_0\tau_D-t) 
\end{equation}
Here $t_0$ can be understood as the initial time the wave packet starts to move.
 
Thus using only the time operator $M_T(t)$ we have got the formula that bounds
the evolution parameter $\tau_D$ with the time $t$. As one can see in this
very simple case they are proportional one to another and if only $\beta_0=1$ 
and $t_0=0$ these two quantities are the same. 

According to the idea of projection evolution the above procedure gives us only 
an information about temporal structure of our state.

Another way is to "rebuild" the detector making use of the "time trigger".
In this case we have to reconstruct the evolution operator
(\ref{eq:III.4}) adding the resolution of unity $M_T(t)$, $t \in R^3$ as 
follows:
\begin{equation} \label{eq:III.4a}
\Mew{\tau}{\nu} = \begin{cases}
U(\tau-\tau_0)\,M_\Wp(\nu)\,U^\dagger(\tau-\tau_0); & 
	\text{for $\tau_0 \leq \tau < \tau_D$ and $\nu$ of the form 
	(\ref{eq:III.1})}, \\
M_D(\nu); & 
	\text{for $\tau = \tau_D$ and $\nu = \Delta$ or $\nu = 0$}, \\
M_T(t); &
	\text{for $\tau = \tau_D+\epsilon$ and $t \in R$},
\end{cases}
\end{equation}
where $\epsilon$ is an arbitrary small positive number which is only formally
needed but all results are independent of it.  

Now the probability (density) $\Prob(\tau_D+\epsilon;\mu,\Delta,t)$  of
the situation that $\rho(\tau_D+\epsilon)$ is the  density operator describing
the system state existing in time $t$  can be calculated multiplying the
conditional probability (\ref{eq:IV.3}) and the probability of finding our
particle in the state (\ref{eq:IV.2}) which is equal to the denominator of the
r.h.s. of (\ref{eq:IV.2}), for details see \cite{Gozd03}.

According to our rules the state $\rho(\tau_D+\epsilon)$ can be written as:
\begin{equation} \label{eq:IV.2a}
\rho(\tau_D+\epsilon;\nu_1=\mu,\nu_2=\Delta) = 
\frac{M_T(t)M_D(\Delta)\,U(\tau_D)\,M_\Wp(\mu)\,\rho_0\, 
      M_\Wp(\mu)\,U^\dagger(\tau_D)\,M_D(\Delta)M_T(t)}
{\Trace [\,M_T(t)M_D(\Delta)\,U(\tau_D)\,M_\Wp(\mu)\,\rho_0\, 
	   M_\Wp(\mu)\,U^\dagger(\tau_D)\,M_D(\Delta)M_T(t)\,]}.
\end{equation}
The trace in the denominator is the required probability
$\Prob(\tau_D+\epsilon;\mu,\Delta,t)$ of a particle detection  after
the etime $\tau_D$  at time $t$. Making use of the equation (\ref{eq:IV.3})
the probability can be easily expressed as:
\begin{equation}\label{eq:IV.3a}
\Prob(\tau_D+\epsilon;\mu,\Delta,t)=
 	\int_\Delta d^3 \vec{x}\ \left|\,\bra{t,\vec{x}}\,U(\tau_D)\,
	\ket{\mu} \right|^{\;2}
\end{equation}
which after separating  space and time dependent factors can be further
rewritten as:
\begin{equation}\label{eq:IV.3b}
\Prob(\tau_D+\epsilon;\mu,\Delta,t)=
\left|\,\bra{t}\,e^{i\beta_0 \hat{k}^0\tau_D}\,\ket{\mu^0} \right|^{\;2}
\int_\Delta d^3 \vec{x}\ \left|\,
        \bra{\vec{x}}\,
         e^{-i\beta\hat{\vec{k}}^2\tau} \,
	\ket{\vec{\mu}} \right|^{\;2},
\end{equation}
where $\ket{\vec{\mu}}$ denotes the spatial part of the wave packet.

We see that, under the same assumptions as for "information observable" case we
have obtained again a $\delta$ type dependence in the numerator of 
(\ref{eq:IV.3b}). The time dependence is exactly the same as in  previous case.
This makes the  analysis easier and more instructive. \\ 
Of course, this is not the rule and in general the dependence $\tau
\leftrightarrow t$ can be much more complicated.

\section{Discussion on the detection moment probability}
\label{sec:results}
One of the  advantages of the projection evolution scheme is that there
is a possibility to estimate the moment of occurring the measurement. The
formula (\ref{eq:III.7}) gives the probability of the particle detection after
the specified evolution steps $\tau_D$. We have assumed the particle is
represented by a wave packet of the form (\ref{eq:III.1}) with (\ref{eq:III.9}),
which corresponds to the particle moving with the average velocity $\vec{v}_g$
as introduced in (\ref{eq:III.15}).

\begin{figure}[ht]
\begin{center}
\includegraphics[angle=-90,width=10cm]{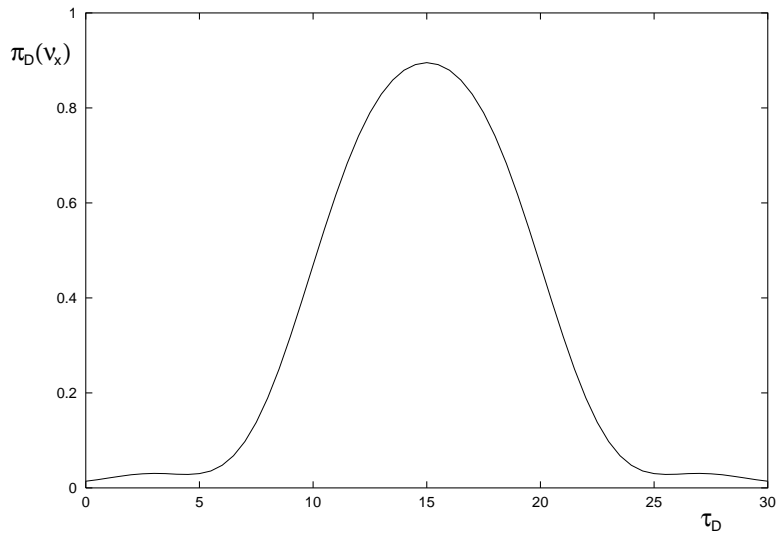}
\caption{\label{fg:1}Particle detection probability with respect to 
the final etime $\tau_D$.}
\end{center}
\end{figure}

Let us consider as an example the electron moving towards the cubic detector
standing some distance far from the initial particle location. To simplify
considerations we have calculated the probability of the particle detection in
one dimension only, choosing the axis parallel to the electron linear momentum
vector. 
The figure (\ref{fg:1}) presents the particle detection probability calculated
using (\ref{eq:III.15}) for several various final etimes $\tau_D$.

The results, of course, depend on some essential system properties, like the
particle speed and the region $\Delta$ of the space occupied by the detector.
However, from this example one can draw quite general conclusions. \\  In the
figure, the source of the particle is at the origin of the coordinate system.
The group velocity of the wave packet  $v_g=0.5$5 and our one dimensional 
detector occupies the region between $x=5$ and $x=10$ on the "$Ox$" axis. All
the physical quantity are in some arbitrary units.

As one can see, there is a non-zero probability that the particle will be
detected at the beginning of its evolution and that probability raises until it
achieves the maximum value at the etime $\tau_D=15$. According to the previous
section we can bound the evolution step parameter $\tau_D$ with the real time
$t$, which in the simplest case are equal one to another. 
In that way the maximum of the detection probability corresponds to the 
classical time $t=s/{v_g}$, where $s$ is the  distance between the source of 
the  particles and the  middle point of the detector $x_D=7.5$. 

The figure suggests that the particle will be detected with large
probability if only it touches the detection area but there is also possible it
will not be registered at all during its evolution.

Further analysis shows that the detection probability is strongly dependent on
the detector shape and the uncertainty of the particle localization. Especially
in case of well defined linear momentum of the particle ($\Delta k\approx0$)
its localization is spread over a large region of the space and if  the
detector does not cover a sufficient piece of the space the particle can omit it.

\begin{figure}[ht]
\begin{center}
\includegraphics[angle=-90,width=15cm]{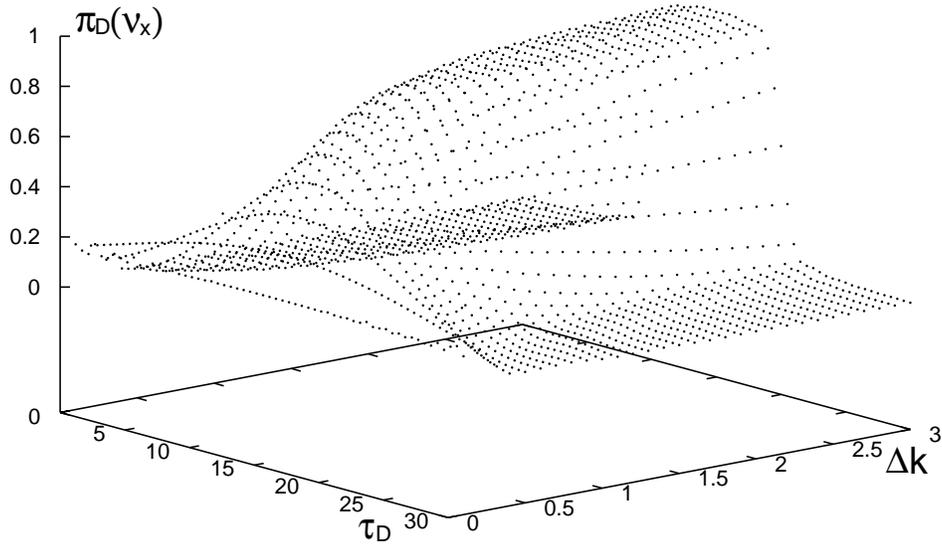}
\caption{\label{fg:2}Particle detection probability with respect to the 
evolution steps $\tau_D$ and the wave packet width $\Delta k$.}
\end{center}
\end{figure}

The figure (\ref{fg:2}) shows the particle detection probability for specified
$\tau_D$ with respect to the wave packet width $\Delta{k}$, that is for various
values of the localization and linear momentum uncertainty.
As one can see if the linear momentum of the electron is well known, i.e.
$\Delta k\approx0$ and the position is not well determined,
detection of the particle is very difficult -- almost impossible. The more
precise are both the quantities, that is when minimizing the uncertainty due to
the Heisenberg uncertainty principle, the more sharp and effective is the
measurement; that means the probability of the measurement is almost equal 0
outside the detector and nearly equal 1 for the time corresponding to the
detection area arrival. \\
Although the linear approximation (\ref{eq:III.12}) do not allow us to use
(\ref{eq:III.15}) when the distribution of linear momentum is not
narrow,  some numerical calculations shows that in that case the particle
detection has small probability with the maximum value not greater than 0.2.
This means the particle will come across the detector and probably will not be
registered. 
\begin{figure}[ht]
\begin{center}
\includegraphics[angle=-90,width=10cm]{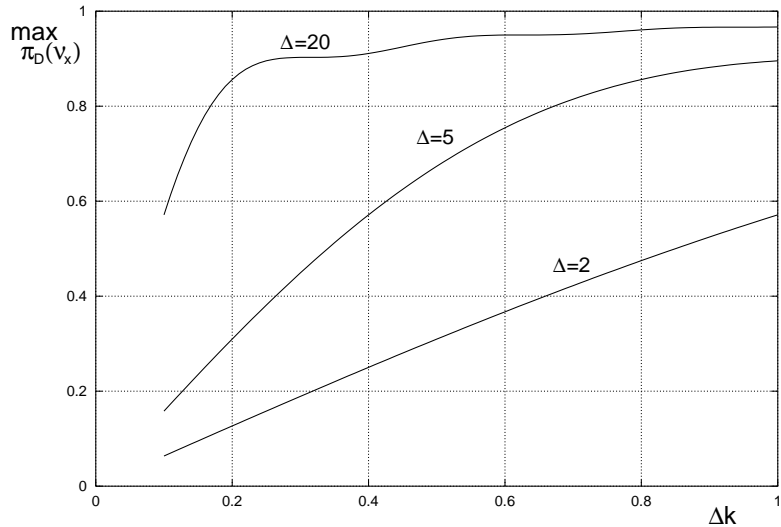}
\caption{\label{fg:3}The maximum of the particle detection probability 
with respect to the wave packet width $\Delta k$ and the detector 
length\;$\Delta$.}
\end{center}
\end{figure}
According to the figure (\ref{fg:3}) one can find that the maximum of the
measurement probability depends on the detector shape, i.e. the more larger
is the detector the more probably it registers the particles, even if they are
not well localized in space ($\Delta k \approx 0.2$). The detector can scan
larger area and more wave packet is  overlapped.

The problem is very similar if we consider the wave packet described by the
coherent state, described i.e. by (\ref{eq:III.1}) with (\ref{eq:III.9}) and 
\begin {displaymath}
\alpha_\mu (\vec k)=\bigg(\frac{\sigma}{\pi}\bigg)^{3/4}
e^{-\frac{\sigma {(\vec K-\vec k)}^2}{2}}
e^{i(\vec K-\vec k)a}
\end {displaymath}
where $\vec K$ denotes the linear momentum vector of the particle, $a$ is the
initial particle position and $\sigma$ denotes the wave packet width. The
detection probability, however, is in this case very alike the presented 
in the figure (\ref{fg:1}).
There is no qualitative differences in the evolution or measurement
description, only the calculations are more difficult and have to be made using
some numerical recipes.

\begin{figure}[ht]
\begin{center}
\includegraphics[angle=-90,width=10cm]{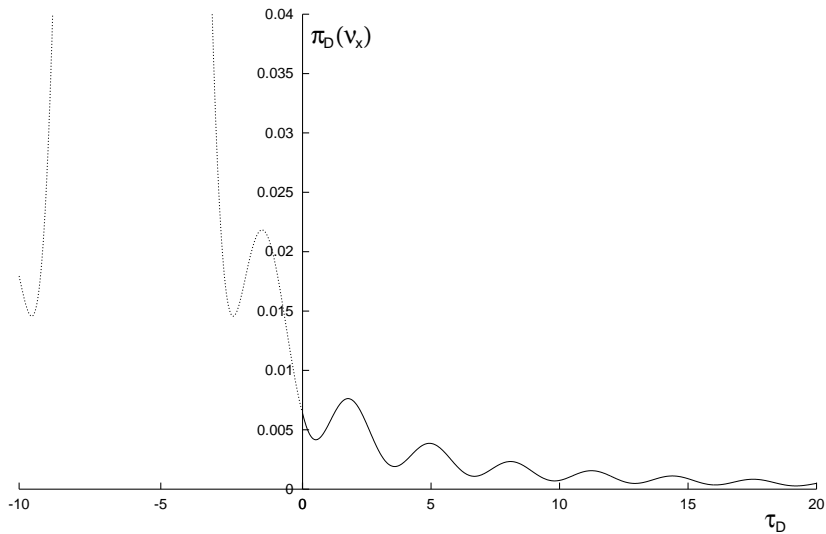}
\caption{\label{fg:4}Particle detection probability with respect to 
the etime $\tau_D$ in case of moving in the opposite direction.}
\end{center}
\end{figure}

Another interesting situation is when the particle is moving not toward the
detector, but in quite opposite direction. The probability of the
detection, assuming the same parameters as previously, but with the group
velocity $\vec{v'}_g=-\vec{v}_g$, is presented on the figure (\ref{fg:4}).
The far the particle is from the detection area the lower is the probability of
its detection, and the probability, even if exist is very small, nearly equal
zero. That means there is a little chance to detect such a particle even if it
is  far from the measurement device.

\section{Summary}
\label{sec:summary}
In this paper we have presented a problem of  time relations for 
the particle detection process. A
closed system of the particle moving in the neighborhood of the measurement
device has been described in terms of projection evolution, according to the
postulated new Law of Nature. The example shows how to introduce in quantum
mechanics the notion of time as an observable.  

The idea gives an  opportunity to
build a time operator as the observable projecting on the subspace of
simultaneous events. Thus we are able to obtain the probability of the particle
measurement as a function of time which was not possible within the standard
quantum mechanics.

Some calculations within a schematic model has been presented.  The results are
very close to our intuition about the process giving the proper dependences
which comparable  with the experiments. 

It is also important to note that the projection evolution postulate let us to
describe the decoherence as the inner property of quantum objects.

\end{document}